# Two Dimensions for Organizing Immersive Analytics: Toward a Taxonomy for Facet and Position

**David Saffo**
Northeastern University
Boston, MA 02115, USA
saffo.d@husky.neu.edu

**Sara Di Bartolomeo**
Northeastern University
Boston, MA 02115, USA
dibartolomeo.s@husky.neu.edu

**Caglar Yildirim**
Northeastern University
Boston, MA 02115, USA
c.yildirim@northeastern.edu

**Cody Dunne**
Northeastern University
Boston, MA 02115, USA
c.dunne@northeastern.edu

## Abstract
As immersive analytics continues to grow as a discipline, so too should its underlying methodological support. Taxonomies play an important role for information visualization and human computer interaction. They provide an organization of the techniques used in a particular domain that better enable researchers to describe their work, discover existing methods, and identify gaps in the literature. Existing taxonomies in related fields do not capture or describe the unique paradigms employed in immersive analytics. We conceptualize a taxonomy that organizes immersive analytics according to two dimensions: spatial and visual presentation. Each intersection of this taxonomy represents a unique design paradigm which, when thoroughly explored, can aid in the design and research of new immersive analytic applications.

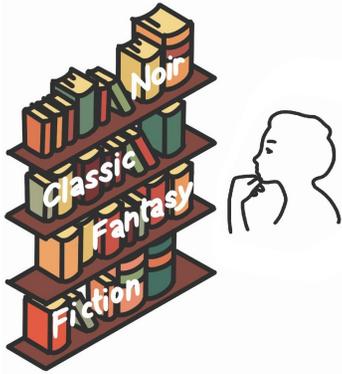

**Figure 1:** A user looks at the experiment presented by Bach et al. in [3], in which an augmented reality layer is placed over a library to display information about the books. An example of a medium-sized situated visualization.

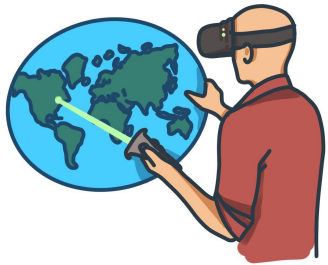

**Figure 2:** A user explores a world map that can be moved around the virtual space, as in the experiment by Yang et al. [14]. This is an example of medium fixed global position.

## Introduction

Conceptual models, frameworks, typologies, and taxonomies are central to the canon for every research area. These works provide researchers with methodological support for pedagogy, evaluation, and novel research. Immersive Analytics (IA) is in its infancy and, as a result, many of these methods have yet to be fully proposed, accepted, and validated. IA researchers have turned to adjacent research areas (HCI, InfoVis, VR/MR) for methodological support. But IA poses unique challenges, aspects, and applications that are not fully represented in other fields. Recent work has tried to address these issues by extending existing methodological frameworks to best fit IA. In this paper we propose a taxonomy that aims to organize the design space of IA facet and position according to two dimensions preeminent in IA design: **spatial and visual presentation**.

Defining methods for spatial and visual presentation provides researchers with a high-level overview of possible approaches. Detailing the intersections of these methods provides the specific encodings, interactions, and applications. With this knowledge, researchers that have chosen their approach for facet and position can select the appropriate techniques for building their respective IA application, identify the need for novel methods, more easily categorize their work, or simply study the field as it evolves.

This position paper serves as an initial spatial and visual presentation taxonomy. Herein we describe our reasoning for the selected dimensions and demonstrate how they can be used. Our goal is to start a community discussion on how to best evolve this taxonomy to support IA research.

## Related Work

IA research draws most of its methodological support from information visualization. These include general purpose methods such as Munzer's nested model [11], as well as specialized works as in Bach's et al. trajectory descriptive framework [1]. These general methods are useful for IA and their influence can be seen in much of IA research. However, general methods leave many questions that IA researchers they will need to answer on their own if they wish to successfully co-opt them for IA.

Researchers have addressed this challenge by extending existing literature to better suit IA. Marriott et al. [10] extend Brehmer's multi-level task topology [4] by adding two more levels and additional parameters specific to IA. They detail 5 levels of this framework consisting of where, who, what, why, and how — where and who being the added levels. This framework shows promise for guiding IA researchers towards effective design considerations. However, discovering literature that describes the outcomes of these considerations is difficult. While the IA design space is massive and continuously evolving with new technology and paradigms, it is currently lacking in methodological support for defining, organizing, and detailing this design space.

## Facet and Position Taxonomy

Facet and position are terms used in IA to encompass how to arrange views or elements relative to each other, viewers, or even the real world [10]. It is our belief that facet and position are what differentiates the design of immersive analytics the most from traditional visual analytics and information visualization. The design space of facet and position for traditional information visualization is well defined with 2D displays, limiting the possible arrangement of views. For example, presenting a bar chart on a 2D display is relatively straightforward, while IA can have many more possibilities that haven't all been explored. Position and size will affect how the user experiences the visualization, and can be used to convey a meaning. Presenting a bar chart in this context is a considerably more delicate task for designers, because they have to choose where the bar chart should be positioned and scaled in infinite 3D space. Hence, the design space for facet and position for IA is vast and complex to navigate, and there are no clear design recommendations for position, orientation, scale, and presentation of views. Marriott et al. [10] use the terms facet and position together to describe collectively the arrangement of different views and their placement in the world, but we believe it should be further sliced in two more dimensions: **visual presentation** — what views the user will see — and **spatial presentation** — how the user will experience views.

**Spatial Presentation:** Immersive 3D views are presented in virtual or mixed space that can contain an infinite canvas, virtual room, or other augmented views. We describe spatial presentation in terms of how this space is presented to the user. The *presentation* will dictate the context and location of where views will be seen and how they will be interacted with. While not an exhaustive list, we postulate the following categories of spatial presentation: free global position, fixed global position, and situated. **Free global position** constitutes a spatial presentation in which a user can actively move or be moved within, such as presenting a room in which views are arranged and users can move about freely. Alternatively, **fixed global position** dictates a spatial presentation in which users are fixed into one predetermined location, such as views where the user will be still in the virtual world while using interaction to move the view and elements around them. Lastly, **situated** spatial presentation will place views into a real world setting where the space itself nor the users position can be tangibly manipulated by the application.

**Visual Presentation:** Visual presentation encapsulates how views are positioned, scaled, orientated, and manipulated. These parameters influence what the user will see and how they will analyze and interact with the view. For the sake of conciseness we will only discuss one of these parameters for this version of the taxonomy — scale. While it is possible to define scale on visualizations of real objects, relative to the real size, it is not possible to do the same with abstract visualization. To be consistent within the two categories, scale is defined relative to the average size of a human being and not relative to the original size of the represented item. A view with a **small** scale is one that would reasonably fit into one or two hands. This could be a model of a real world object, such as a rocket, that is shrunk down allowing a user to view more of the model from a single view as well as easily manipulate it in three dimensions. Scaling up from there, a **medium** scale would roughly constitute a view that exceeds no more than the size of an average human being. An example of this could include an abstract visualization, such as a node-link diagram, that is not too small or too large for summative analysis. Lastly, a **large** scale includes everything larger than an average human being. These views would require a user's position or the view's position to change in order to experience the visualization in its entirety. For example, a human organ that can be scaled up significantly in order for users to view its internal structure up close.

*The Taxonomy*
We introduce this prototype taxonomy organized by the two previously described dimensions: visual presentation and spatial presentation (Table 1). At each cell of the grid there is a unique set of methods and techniques for interaction, multi-sensory presentation, and encoding. We see the usefulness of this taxonomy in allowing researchers to:

- Explore the work that has already been done in a specific context to learn about the field,

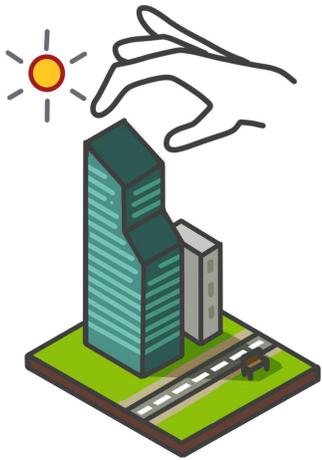

**Figure 3:** A simplified representation of the architectural planning example

|  | **Free Global Position** User can move in virtual world. Virtual world can be moved. | **Fixed Global Position** User is still in virtual world. World can be moved. | **Situated** Position of user and world depend on real world |
|---|---|---|---|
| **Small** Can fit in the palm of a hand | 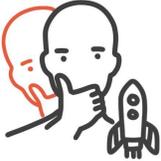 User rotates her head around a small representation of a rocket  Example: [6] | 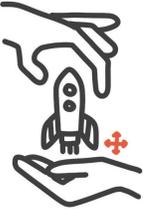 A small visualization being translated and rotated by hands  Example: [9] | 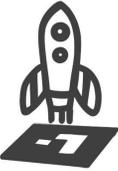 The visualization is on top of an AR Marker, located in the real world  Example: [2] |
| **Medium** Human sized | 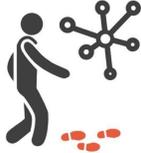 User walks around a medium sized representation of a network  Example: [5] | 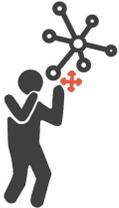 User translates and rotates a medium sized representation of a network.  Example: [7] | 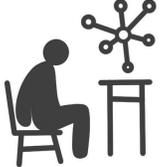 The network visualization is presented on top of a table present in the real world  Example: [13] |
| **Large** Can be walked in | 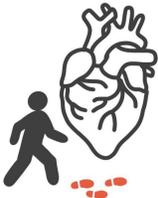 User walks around a large representation of a heart  Example: [12] | 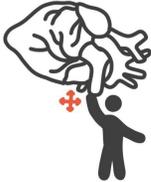 User's position is fixed while manipulating a large heart  Example: [8] | 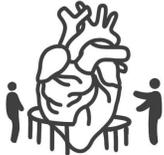 Two users discuss a large visualization of a heart located on a pedestal  Example: [3] |

**Table 1:** Table representation of the proposed taxonomy. Columns indicate position, rows indicate space.

| | |
|---|---|
| **Where:** | HMD VR, using VR controllers for input. World knowledge to the system is given by head and controller tracking. |
| **What:** | 3D model of the buildings and amenities surrounding a specific location. Additional information is given by a simulation of the influences generated by target building to its surroundings (e.g. shadows). |
| **Who:** | Architect (domain expert). Collaboration is not supported |
| **Why:** | High-level tasks: Discover effect of the placement of a target building. Medium-level tasks: explore, compare between different possible solutions |
| **How:** | Digital overlay of 3D data on a 3D model of the environment. Can orient the model at any angle. Can change variables about the environment (e.g. position of the sun) and the target building (e.g. shape of the building). Fixed global position, small. |

**Figure 4:** The 5 questions framework used on the architectural planning example

- Find and reason about design decisions when they know which intersection they are working in, and
- Find gaps in the techniques for an intersection.

Each intersection in the matrix represents a unique design paradigm with its respective literature and applications. These works detail the methods, techniques, interactions, and encodings used in each paradigm. Each cell will also contain a description of each context. Having two broad dimensions allows the taxonomy to be easily extended to include new rows and columns when the need arises. Indeed, as new paradigms form or technologies develop, they can be added to the taxonomy accordingly.

## Using the design space

This taxonomy will work best in conjunction with the 5 question framework presented by Marriott et al. [10]. After answering the first 4 questions (who, where, what, why) researchers can reason about which intersection(s) of the taxonomy will define facet and position. Then, using the literature described by their selected paradigm(s) researchers can more effectively determine how to answer the fifth question — how. We demonstrate how this taxonomy could benefit IA researchers through a design case study.

*Case Study: Architectural planning*
A visualization designer wants to implement a simulation to visualize the effect that the placement of a skyscraper would have on its surroundings. In historical capitals, it is important to take into account how a skyscraper affects the skyline of the city and many cities have regulations in place to avoid buildings from covering, for example, the view of a monument. The intended user, an architect, is interested in studying a skyscraper design immersed in a specific context. The visualization, shown in Figure 3, presents a 3D model of the building and its surroundings. The architect wants to explore the model while sitting at their desk and wants to change some variables such as the position of the sun to study how the building will produce shadows. There will be no collaborators. Since the user wants to explore the model on her desk, the visualization offers a small model that can be comfortably rotated in any direction. In table 1, the cell "fixed global position small" is the one that best describes this situation. An example of the 5 questions framework applied to this case can be found in table 4.

## Future Work and Conclusion

The topics and discussions herein are not meant to be definitive, but rather a call for collaboration to better define those aspects that are particular to the context of Immersive Analytics. We encourage the community to keep up the discussion towards establishing this design space. Moving forward, we hope to focus on validating the taxonomy in conjunction with the 5 questions approach [10] and to refine the cells in the taxonomy. We plan to release an interactive web version of the taxonomy where users can browse and add to the cells, explore a detailed collection of examples, and see a more specific descriptions. As the design space grows with new paradigms and literature, we can began to learn and understand the encodings, interactions, and applications that comprise immersive analytics.

We introduced our perspective on two significant factors for Immersive Analytics: spatial and visual presentation. These factors are critical to understanding and defying facet and position for IA design. We propose a preliminary taxonomy for incorporating these elements into an IA design space. As we believe a domain-defining taxonomy can only be achieved with the help of the community, we ask for collaboration to further this work.